\documentstyle[11pt]{article}

\def\eqa{\begin{eqnarray}}
\def\eqae{\end{eqnarray}}

\def\half{\frac{1}{2}}
\def\fracm#1#2{\hbox{\large{${\frac{{#1}}{{#2}}}$}}}
\def\@magscale#1{ scaled \magstep #1}

\catcode`@=11  
\def\un#1{\relax\ifmmode\@@underline#1\else
        $\@@underline{\hbox{#1}}$\relax\fi}
\catcode`@=12

\let\slo=\o                     

\def\a{\alpha}
\def\b{\beta}

\def\d{\delta}
\def\e{\epsilon}

\def\g{\gamma}

\def\l{\lambda}
\def\m{\mu}
\def\n{\nu}
\def\o{\omega}
\def\p{\pi}
\def\q{\theta}
\def\r{\rho}
\def\s{\sigma}
\def\t{\tau}

\def\z{\zeta}

\def\G{\Gamma}

\def\L{\Lambda}
\def\O{\Omega}

\def\S{\Sigma}
\def\U{\Upsilon}

\def\dslash{\not{\hbox{\kern-2pt $\partial$}}}
\def\Dslash{\not{\hbox{\kern-4pt $D$}}}
\def\pslash{\not{\hbox{\kern-2.3pt $p$}}}
 \newtoks\slashfraction
 \slashfraction={.13}
 \def\slash#1{\setbox0\hbox{$ #1 $}
 \setbox0\hbox to \the\slashfraction\wd0{\hss \box0}/\box0 }
 
\font\ro=cmsy10                          
\def\kcr{{\hbox{\ro \char'170}}}                
\def\ktl{{\hbox{\ro \char'170}}}        
\def\ktr{{\hbox{\ro \char'170}}}        
\def\kbl{{\hbox{\ro \char'170}}}        
\def\kbr{{\hbox{\ro \char'170}}}        

\def\plpl{\raise-2pt\hbox{$\raise3pt\hbox{$_+$}\hskip-6.67pt\raise0.0pt
\hbox{$^+$}\hskip 0.01pt$}}
\def\mimi{\raise-2pt\hbox{$\raise3pt\hbox{$_-$}\hskip-6.67pt\raise0.0pt
\hbox{$^-$}\hskip 0.01pt$}} 

\def\bo{{\raise.15ex\hbox{\large$\Box$}}}               
\def\pa{\partial}                                       
\def\TH{{\raise.2ex\hbox{$\displaystyle \bigodot$}\mskip-4.7mu \llap H
\;}}
\def\face{{\raise.2ex\hbox{$\displaystyle \bigodot$}\mskip-2.2mu \llap
{$\ddot
        \smile$}}}                                      

\def\leftrightarrowfill{$\mathsurround=0pt \mathord\leftarrow \mkern-6mu
        \cleaders\hbox{$\mkern-2mu \mathord- \mkern-2mu$}\hfill
        \mkern-6mu \mathord\rightarrow$}
\def\dvec#1{\vbox{\ialign{##\crcr
        \leftrightarrowfill\crcr\noalign{\kern-1pt\nointerlineskip}
        $\hfil\displaystyle{#1}\hfil$\crcr}}}           

\def\fracm#1#2{\hbox{\large{${\frac{{#1}}{{#2}}}$}}}
\def\frac#1#2{{\textstyle{#1\over\vphantom2\smash{\raise.20ex
        \hbox{$\scriptstyle{#2}$}}}}}                   
\def\sfrac#1#2{{\vphantom1\smash{\lower.5ex\hbox{\small$#1$}}\over
        \vphantom1\smash{\raise.4ex\hbox{\small$#2$}}}} 
\def\bfrac#1#2{{\vphantom1\smash{\lower.5ex\hbox{$#1$}}\over
        \vphantom1\smash{\raise.3ex\hbox{$#2$}}}}       
\def\afrac#1#2{{\vphantom1\smash{\lower.5ex\hbox{$#1$}}\over#2}}    

\newskip\humongous \humongous=0pt plus 1000pt minus 1000pt
\def\caja{\mathsurround=0pt}
\def\eqalign#1{\,\vcenter{\openup2\jot \caja
        \ialign{\strut \hfil$\displaystyle{##}$&$
        \displaystyle{{}##}$\hfil\crcr#1\crcr}}\,}
\newif\ifdtup

\parskip=\medskipamount                 
\thicklines                         

\def\oldheadpic{                                
        \setlength{\unitlength}{.4mm}
        \thinlines
        \par
        \begin{picture}(349,16)
        \put(325,16){\line(1,0){4}}
        \put(330,16){\line(1,0){4}}
        \put(340,16){\line(1,0){4}}
        \put(335,0){\line(1,0){4}}
        \put(340,0){\line(1,0){4}}
        \put(345,0){\line(1,0){4}}
        \put(329,0){\line(0,1){16}}
        \put(330,0){\line(0,1){16}}
        \put(339,0){\line(0,1){16}}
        \put(340,0){\line(0,1){16}}
        \put(344,0){\line(0,1){16}}
        \put(345,0){\line(0,1){16}}
        \put(329,16){\oval(8,32)[bl]}
        \put(330,16){\oval(8,32)[br]}
        \put(339,0){\oval(8,32)[tl]}
        \put(345,0){\oval(8,32)[tr]}
        \end{picture}
        \par
        \thicklines
        \vskip.2in}
\def\oldtitle#1#2#3#4{\oldheadpic\begin{center}\vglue.5in{\large\bf
#1}\\[.6in]
        {#2}\\[.1in] {\it Department of Physics and Astronomy}\\
        {\it University of Maryland, College Park, MD 20742}\\[.6in]
        Physics Publication \#{#3}\\ {#4}\\[1.5in] {\bf
ABSTRACT}\\[.1in]
        \end{center} \begin{quotation}}                 
\def\oldTitle#1#2#3#4#5#6#7{\oldheadpic\begin{center} \vglue .4in
        {\large\bf #1}\\[.4in]
        {#2}\\[.1in] {\it Department of Physics and Astronomy}\\
        {\it University of Maryland, College Park, MD 20742}\\[.1in]
        {#3}\\[.1in] {\it {#4}}\\ {\it {#5}}\\[.4in]
        Physics Publication \#{#6}\\ {#7}\\[.5in] {\bf ABSTRACT}\\[.1in]
        \end{center} \begin{quotation}}                 
\def\border{                                            
        \setlength{\unitlength}{1mm}
        \newcount\xco
        \newcount\yco
        \xco=-21
        \yco=12
        \begin{picture}(140,0)
        \put(\xco,\yco){$\ktl$}
        \advance\yco by-1
        {\loop
        \put(\xco,\yco){$\kcr$}
        \advance\yco by-2
        \ifnum\yco>-240
        \repeat
        \put(\xco,\yco){$\kbl$}}
        \xco=158
        \yco=12
        \put(\xco,\yco){$\ktr$}
        \advance\yco by-1
        {\loop
        \put(\xco,\yco){$\kcr$}
        \advance\yco by-2
        \ifnum\yco>-240
        \repeat
        \put(\xco,\yco){$\kbr$}}
        \put(-20,13){\tiny University of Maryland Elementary Particle
        Physics University of Maryland Elementary Particle Physics
        University of Maryland Elementary Particle Physics}
        \put(-20,-241.5){\tiny  The University of Iowa Particle Theory
        Group The University of Iowa Particle Theory Group The
        University of Iowa Particle Theory Group The University}
        \end{picture}
        \par\vskip-8mm}
\def\bordero{                                           
        \setlength{\unitlength}{1mm}
        \newcount\xco
        \newcount\yco
        \xco=-31
        \yco=12
        \begin{picture}(140,0)
        \put(\xco,\yco){$\ktl$}
        \advance\yco by-1
        {\loop
        \put(\xco,\yco){$\kclr}
        \advance\yco by-2
        \ifnum\yco>-240
        \repeat
        \put(\xco,\yco){$\kbl$}}
        \xco=151
        \yco=12
        \put(\xco,\yco){$\ktr$}
        \advance\yco by-1
        {\loop
        \put(\xco,\yco){$\kcr$}
        \advance\yco by-2
        \ifnum\yco>-240
        \repeat
        \put(\xco,\yco){$\kbr$}}
        \put(-20,12){\ooobacdefghidfghghdhededbihdgdfdfhhdheidhdhebaaahjhhdahbahgdedgehgfdiehhgdigicba}
        \put(-20,-241.5){\oooababaighefdbfghgeahgdfgafagihdidihiidhiagfedhadbfdecdcdfagdcbhaddhbgfchbgfdacfediacbabab}
        \end{picture}
        \par\vskip-8mm}
\def\headpic{                                           
        \indent
        \setlength{\unitlength}{.4mm}
        \thinlines
        \par
        \begin{picture}(29,16)
        \put(165,16){\line(1,0){4}}
        \put(170,16){\line(1,0){4}}
        \put(180,16){\line(1,0){4}}
        \put(175,0){\line(1,0){4}}
        \put(180,0){\line(1,0){4}}
        \put(185,0){\line(1,0){4}}
        \put(169,0){\line(0,1){16}}
        \put(170,0){\line(0,1){16}}
        \put(179,0){\line(0,1){16}}
        \put(180,0){\line(0,1){16}}
        \put(184,0){\line(0,1){16}}
        \put(185,0){\line(0,1){16}}
        \put(169,16){\oval(8,32)[bl]}
        \put(170,16){\oval(8,32)[br]}
        \put(179,0){\oval(8,32)[tl]}
        \put(185,0){\oval(8,32)[tr]}
        \end{picture}
        \par\vskip-6.5mm
        \thicklines}
\def\title#1#2#3#4{\border\headpic {\hbox to\hsize{#4 \hfill UMDEPP#3}}\par
        \begin{center} \vglue .5in {\large\bf #1}\\[.6in]
        {#2}\\[.1in] {\it Department of Physics and Astronomy}\\
        {\it University of Maryland, College Park, MD 20742}\\[1.5in]
        {\bf ABSTRACT}\\[.1in] \end{center} \begin{quotation}}  
\def\Title#1#2#3#4#5#6#7{\border\headpic
        {\hbox to\hsize{#7 \hfill UMDEPP #6}}\par
        \begin{center} \vglue .4in {\large\bf #1}\\[.4in]
        {#2}\\[.1in] {\it Department of Physics and Astronomy}\\
        {\it University of Maryland, College Park, MD 20742}\\[.1in]
        {#3}\\[.1in] {\it {#4}}\\ {\it {#5}}\\[.5in] {\bf ABSTRACT}\\[.1in]
        \end{center} \begin{quotation}}                 
\def\endtitle{\end{quotation}\newpage}                  

\def\ad{{\kern0.5pt \alpha \kern-5.05pt \raise5.8pt\hbox{$\textstyle.$}\kern0.5pt}}

\def\bd{{\kern0.5pt   \beta \kern-5.05pt \raise5.8pt\hbox{$\textstyle.$}\kern0.5pt}}

\def\qd{{\kern0.5pt q \kern-5.05pt \raise5.8pt\hbox{$\textstyle.$}\kern0.5pt}}
\def\Dot#1{{\kern0.5pt {#1} \kern-5.05pt
\raise5.8pt\hbox{$\textstyle.$}\kern0.5pt}}

\begin{document}

\def\gfrac#1#2{\frac {\scriptstyle{#1}}
        {\mbox{\raisebox{-.6ex}{$\scriptstyle{#2}$}}}}
\def\gg{{\hbox{\sc g}}}
\border\headpic {\hbox to\hsize{Jnauary 2002 \hfill{UMDEPP 02-028}}}
\par
{\hbox to\hsize{$~$ \hfill
{UIOWA 02-001}}}
\par
{\hbox to\hsize{$~$ \hfill
{CALT-68-2367}}}
\par
\setlength{\oddsidemargin}{0.3in}
\setlength{\evensidemargin}{-0.3in}
\begin{center}
\vglue .15in
{\Large\bf Chiral Supergravitons Interacting with a 0-Brane
$N$-Extended NSR Super-Virasoro Group\footnote{
Supported in part by National Science Foundation Grants 
PHY-01-5-23911} }
\\[.5in]
A.\ Boveia{\footnote{ Present Address: University of California at
    Santa Barbara, USA}}, Bj\slo rg A.\ Larson, and V.\ G.\ J.\
Rodgers{\footnote{vincent-rodgers@uiowa.edu}}
\\[0.06in]
{\it Department of Physics and Astronomy\\
The University of Iowa\\ 
Iowa City, IA 52246-1479 USA}\\[.05in]
~\\
S.\ James Gates, Jr.\footnote{On Sabbatical leave at
the California Inst. of Technology, Sept. 2001 thru 
July 2002}${}^,$\footnote{gatess@wam.umd.edu}, W.\ D.\ 
Linch, III\footnote{linch@bouchet.physics.umd.edu}, and 
J.\ A.\ Phillips\footnote{ferrigno@physics.umd.edu}
\\[0.06in]
{\it Department of Physics, University of Maryland\\ 
College Park, MD 20742-4111 USA}\\[.05in]
{\it and }\\
Dagny M. Kimberly\\[0.06in]
{\it Department of Physics, Brown University\\
Providence, RI 02912 USA}\\[.1in]

{\bf ABSTRACT}\\
\end{center}

\begin{quotation}
{We continue the development of ${\cal S}_{AFF}$ by examining the
cases  where there are $N$ fermionic degrees of freedom associated
with a 0-brane.  These actions correspond to the interaction of the
$N$-extended super Virasoro algebra with the supergraviton and the
associated SO($N$) gauge field that accompanies the supermultiplet.  
The superfield formalism is used throughout so that supersymmetry is
explicit.}\\[.2in]
${~~~}$ \newline

\endtitle
\noindent

\section{Introduction}
~~~~In the literature there has been some focus on the role that 
diffeomorphisms play in the development of both classical and quantum
gravity through string theories.  One particular approach uses the
Virasoro algebra and its dual (the coadjoint representation) as a 
guide to these developments \cite{diff1,diff2}.   This approach 
differs from theories of gravity that are based purely on geometry 
since it is the Lie derivative, as opposed to the covariant derivative, 
which is at the center stage. Since the Lie derivative exists in any
dimension, this approach allows one to get a foothold on gravitation 
on the two-dimensional string worldsheet.  From this viewpoint, the
Virasoro algebra corresponds to the time independent  Lie derivative 
of contravariant vector fields, while the isotropy equations on the
coadjoint orbits serve as extensions of the Gauss' Law constraints 
from Yang-Mills theories.  Actions that admit these constraints when
evaluated on two-dimensional surfaces were constructed in
\cite{diff1,diff2}.   Recently \cite{gates-rodgers}, these actions 
were extended to the super Virasoro algebra for one supersymmetry 
($N$ = 1).   Since the two genders of particles, fermions and bosons, 
are placed on equal footing through smooth continuous transformations,
these actions were dubbed {\em affirmative actions}.  

Another way to conceptualize this 
work is to view our problem as an attempt to couple a NSR 0-brane 
with $N$-extended supersymmetry to a background containing 2D, ($N$,0) 
supergravity fields.  This is an extension of the well-known conformal
$\s$-model technique.  There one couples the massless compositions of 
a string theory via a world sheet action to the string.   Conformal
invariance of the world sheet $\s$-model then imposes the equations of
motion on the massless composites.

In our study, we introduce an NSR 0-brane described by coordinates
$\t$ and $\z^{\rm I}$ (I = $1, \, \dots ,\, N$) such that vector 
fields constructed from these coordinates naturally carry an arbitarary
$N$ and model-independent representation of the super-Virasoro algebra.  
These vector fields provide a representation of the centerless
super-Virasoro algebra.  As a benefit of such a geometrical approach, 
the generators of the super-Virasoro algebra naturally possess a
set of transformation laws under the action of the group.  Consequently,
the co-adjoint elements to the generators also naturally obtain a
set of transformation law.   By demanding that these co-adjoint
elements be embedded within background fields and preserve the
symmetries of the co-adjoint elements, restrictions on the backgrounds
are derived.  Thus a major problem for us is to find constructions 
that respect these symmetries within this approach.

The purpose of this work is to show that we can extend these actions  
to $N$ supersymmetries for chiral two dimensional models using the 
results found in Ref.\cite{curto} as a guide.  However, in this work 
we will focus on a superfield formulation as opposed to the component 
construction.  Two superfields are used in the construction, one
corresponds to the $N$-extended prepotentials associated with the
graviton while the other is the prepotential associated with the gauge 
superfield that accompanies the internal SO($N$) symmetry that is now
local.  

\section{$N=1$  Affirmative Actions}
~~~~For the sake of continuity we quickly review some of the results of
Ref.\cite{gates-rodgers}.  Consider the super Virasoro algebra  which
contains the bosonic Virasoro generators  
$L_m, m\,\epsilon\, {\bf Z},$ and fermionic generators $G_\m, \m\, 
\in\, {\bf Z}$ or ${\bf Z} + \frac{1}{2}$, and a central charge
$\hat{c}$.  Its algebra is given by
\begin{eqnarray}
\lbrack L_m, L_n \rbrack &=& (m-n) L_{m+n} + \frac{1}{8} \hat{c} 
(m^3-m) \d_{m+n, \,0} \, I ~~~, \nonumber\\
\lbrack L_m, G_\m \rbrack  &=& (\frac{1}{2} m-\m) G_{m+\m} ~~~, 
\nonumber \\
\{ G_\m, G_\n \} &=& -i \,4 \, L_{\m +\n} -i \frac{1}{2} \hat{c} 
\left( \m^2-\frac{1}{4} \right) \d_{\m + \n , \, 0} \, I ~~~.
 \label{SValg}
\end{eqnarray}
One can show \cite{van} that this algebra can be represented with 
superfields by writing, 
\eqa
 && A (z,\z) = \sum_{m = - \infty}^{\infty} (A^m z^{m+1}) + 2 \z 
\sum_{\m = - \infty}^{\infty} (A^\m  z^{\m + \half}) ~~~,
 \label{SF1} 
\end{eqnarray}
where for now $\z$ is simply a one dimensional Grassmann
coordinate\footnote{We can also refer to this as the NSR 0-brane
Grassmann coordinate.} (later we use $\q$ to denote the space-time
Grassmann supercoordinate variable).  The generic element of the algebra
written in Eq.(\ref{SF1}) has an equivalent representation as a doublet
$(A (Z), a)$ with $Z= \{z,\z \}$.   Then, in terms of derivations on 
the superfield the commutation relations appear as \cite{loops,van}
\begin{equation}
\lbrack [ (A,a) (B,b) ]\rbrack = ( (\partial A) B - A \partial B - i
\half  ({\cal D}A) ({\cal D}B) , \oint dZ (\partial^2 {\cal D}A) \, 
B ) ~~~,
\end{equation}
where $A$ and $B$ are adjoint elements and
\begin{equation}
{\cal D} = \frac{\partial}{\partial \z} + \frac{i}2 \z \frac{\partial}
{\partial z} \qquad, \qquad dZ = \frac{dz}{\, 2\p i} d \z ~~~.
\end{equation}
$ {\cal D}$ implies $ {\cal D}^2 = \frac{i}{2}\pa_z = \fracm 12 
\{  {\cal D}  \, ,\, {\cal D} \}$.

Now we can take an adjoint superfield, say $F$, and act on a coadjoint
element $B^\star$ as a Lie derivative and supersymmetry transformation 
to find \cite{rai,van}
\begin{equation}
\delta_F B^{\star} = -F {\cal D}^2 B^{\star} - \half {\cal D} F 
{\cal D} B^{\star} - \frac{3}{2}{\cal D}^2F B^{\star} + q\,{\cal 
D}^5F ~~~, \label{coad1}
\end{equation}
where $F$ has the decomposition $F=\xi + i \z \epsilon$ and \( B^{\star
}=(u + i \z D,b^{\star}) = (u,D,b^{\star}). \)   The isotropy 
equation (stability equation) for the coadjoint element $B^\star$ is
given by setting Eq.(\ref{coad1}) to zero.  This determines the
subalgebra that will leave the coadjoint vector $B$ invariant.   In 
terms of component fields this becomes the two coupled equations 
\eqa
-\xi \partial D - \half \epsilon \partial u -\frac{3}{2} \partial 
\epsilon u + b^{\star} \partial^3 \xi -2 \partial \xi D =0 
\label{bosonpart}\\
 -\xi \partial u -\half \epsilon D - \frac{3}{2} \partial \xi u + q\, 
\partial^2 \epsilon = 0 \label{fermionpart}
\eqae
with $ \partial = \partial_z $. 
These equations are part of the field equations one extracts from the
affirmative action.  They have a direct analog to the Gauss' Law
constraints found in Yang-Mills.

In order to embed this into two dimensions, we replace the 1D Grassmann 
variable $\z$ with 2-dimensional chiral Majorana spinors $\z^\a$.   Then the 
supersymmetric covariant derivative operator becomes
\begin{equation}
{\cal D}_\m = \partial_\m -\frac i2 \g^N_{\n \m}\z^\n \partial_N  ~~~.
\end{equation}
With this 
\begin{equation}
\{{\cal D}_\m ,{\cal D}_\n \} = -i \g_{\m \n}^N {\partial \over 
\partial z^N} ~~~,
\label{X1}
\end{equation}
where $\g^M_{\mu \nu}$ is the associated Gamma matrix.  We will use
capital Latin indices for space-time indices and small Greek for
spinor indices.  
With these  $\g^M_{\a \b}$'s  we also introduce $\g^{a \a \b}$ such 
that 
\begin{equation}
\g^{A}_{\a \b} \g^{B \b \l} =\frac12 \delta^\l_\a \eta^{A B} +
\frac12 \S^{A B \l}_\a ~~~,
\label{X2}
\end{equation}
where $\S^{A B \l}_\a$ is anti-symmetric in its space-time indices. 
 
Then, for example, an adjoint element is promoted to a vector 
superfield, $F$, and has a $\z$ expansion $F^M =( \xi^M + \z^\a \g_{\a
\b}^M \epsilon^\b  + \sigma \, \z^\m  \z^\a \g^N_{\m \a })$ while a 
two dimensional coadjoint element is promoted to the $\frac32$
spin superfield 
$B_{\m M}=(\U_{\m M} + \z^\a \g_{\a \b}^{N} D_{ M N} + \z^\a 
\z^\b \g^N_{\m [\a } A_{ \b ] M N}) $. 
The transformation  of $B_{\m M}$ with respect to $F^N$ is 
\begin{equation}
\delta_{F} B_{\mu M} = F^N \partial_N B_{\mu M} + \partial_M F^N
B_{\mu N} + \frac12 (\partial_N F^N) B_{\mu M} + i~({\cal D}_\l 
F^N \g^{\l \n}_N){\cal D}_\n B_{\m M}. \label{supercoad}
\end{equation}
This is seen as the Lie derivative with respect to $F^M$ on the space-time 
index in the first three summands followed by a supersymmetry transformation 
on $B_{\m M}$ with $\e^\n \equiv ({\cal D}_\l F^N \g^{\l \n}_N)$. This 
combination of a  Lie derivative and supersymmetry transformation is a 
natural extension of the Lie derivative.  The supersymmetric isotropy 
  equation in higher dimensions can now be thought of as setting
  $\delta_{F} B_{\mu M}$ to zero.  Then those fields $F$ that satisfy
  this condition make up a subspace of vector fields that form the
  isotropy algebra for $B_{\m M}$. 
In order to be consistent with conformal field theory, every spinor
index will carry a density weight of $\frac12$.  Thus the superfield 
$B_{\m N}$ is a tensor density with weight $\frac12$.

\section{Prepotentials for Superdiffeomorphisms and SO($N$)}
\subsection{The Components of the Algebra and Its Dual}
~~~~There have been many $N$-Extended Super Conformal algebras posed in
the literature \cite{Kac}.  However we will focus on the $N$-extended
Super Virasoro algebra proposed in \cite{gates-rana}.  There one has 
$$
\eqalign{ {~~~~}
G_{\cal A} {}^{\rm I} &\equiv~ i \, \t^{{\cal A} + \fracm 12} \, 
\Big[\, \, \pa^{\rm I} ~-~ i \, 2 \, \zeta^{\rm I} \, \pa_{\t} ~ 
\Big] ~+~ 2 (\, {\cal A} \,+\, \fracm 12 \,) \t^{{\cal A} - 
\fracm 12}  \z^{\rm I} \z^{\rm K} \, \pa_{\rm K} ~~~, \cr
L_{\cal A} &\equiv~ -\, \Big[ \, \t^{{\cal A} + 1} \pa_{\t} ~+~ \fracm
12
({\cal A} \, + \, 1) \, \t^{\cal A} \zeta^{\rm I} \, \pa_{\rm I} ~
\Big]  
~~~, \cr
T_{\cal A}^{\, \rm {I \, J }} &\equiv~ \t^{\cal A} \, \Big[ ~ \z^{\rm I} 
\, \pa^{\rm J} ~-~ \z^{\rm J} \, \pa^{\rm I}  ~\Big] ~~~, \cr
U_{\cal A}^{{\rm I}_1 \cdots {\rm I}_q} &\equiv~ i \, (i)^{ [\fracm q2
]} 
\, \t^{({\cal A} - \fracm {(q - 2)}2 )} \zeta^{{\rm I}_1} \cdots \,
\zeta^{{\rm I}_{q-1}} \, \pa^{{\rm I}_q} ~~~,~~~ q \, = \,  3 , \, \dots
\,
, \, N \, + \, 1 ~~~,  \cr
R_{\cal A}^{{\rm I}_1 \cdots {\rm I}_p} &\equiv~  (i)^{ [\fracm p2 
]} \, \t^{({\cal A} - \fracm {(p - 2)}2 )} \zeta^{{
\rm I}_1} \cdots \, \zeta^{{\rm I}_p} \, \pa_{\t} ~~~~~~,~~~ p \, = \, 2
, 
\, \dots \, , \, N  ~~~,  }  \eqno(1)
$$
where  $N$ is the number of supersymmetries.  Here we have used the 
notational conventions of \cite{gates-rana}.  In \cite{curto} one uses the
dual representation of the algebra  in order to develop
a field theory.  This idea is very different from that used in so
called conformal field theories as in this approach there are no fields
external to the algebra (actually its dual) 
introduced,  that is no model or Lagrangian is introduced that is 
 exterior to the algebra.  The impetus behind the construction of the 
 Lagrangian is for its field equations to
 correspond to constraints that arise on the coadjoint orbits.

The elements of the algebra can be realized as fields whose tensor
properties can be determined by they way they transform under 
one dimensional coordinate transformations. How each field transforms
under a Lie derivative with respect to $\xi$  are summarized below. In
conformal field theory these transformation laws characterize the
fields by weight and spin.  Here we treat the algebraic elements as one
dimensional tensors so that higher dimensional realizations can be
achieved.  

\begin{center}
\begin{tabular}{|c|c|c|}\hline
\multicolumn{3}{|c|}{ \bf Table 1: Tensors Associated with the Algebra} \\ \hline\hline
${\rm {Element\, of\, algebra}}$ & ${\rm {Transformation\, Rule}}$ & ${\rm Tensor\, Structure}$ \\ \hline 
$ L_{\cal A} ~\to~ \eta $ &  $ \eta  ~\to~ -\xi' \eta + \xi \eta'  ~~~~$ & $ \eta^a$  \\ \hline
$ G^{\rm I}_{\cal A} ~\to~ \chi^{\rm I} $ &  $ \chi^{\rm
  I}~\to~{-\xi(\chi^{\rm I})' +\fracm 12 \xi' \chi^{\rm I}} ~~ $
&$\chi^{\rm I; \a}$  \\ \hline
$T^{\rm{ R} { S}}~\to~t^{\rm{ R}{ S}} $ &  $ t^{\rm{ R}{ S}}  ~\to~  -\xi\,({t}^{\rm{ R}{S}})'~~~~$ & $ t^{\rm{ R}{ S}} $  \\ \hline
$U^{\rm V_1 \cdots V_n} ~\to~ w^{\rm V_1 \cdots V_n}  $ &  $ w^{\rm
  V_1  \cdots  V_n}\rightarrow  -\xi (w^{\rm V_1
  \cdots  V_n})'  -\fracm 12 (n-2) \xi' w^{\rm V_1 \cdots V_n}~~~~$ &
$w^{\rm V_1 \cdots  V_n;\, a}_{\,\,\,\,\,\,\,\,\,\,\,\, \a_1 \cdots \a_n}  $  \\ \hline
$R^{\rm T_1 \cdots T_n}_{\cal A}~\to~ r^{\rm T_1 \cdots T_n} $ &
${r}^{\rm T_1 \cdots T_n} ~\to~   -({r}^{\rm T_1 \cdots T_n})'\xi-
\fracm 12 (n-2)  \xi'  r^{\rm T_1 \cdots T_n}  ~~~~$ & $ r^{{\rm
    T_1 \cdots T_n}\, ;a }_{\,\,\,\,\,\,\,\,\,\,\,\,  \a_1 \cdots \a_n} $ \\ \hline
\end{tabular}
\end{center}
In the above table we have used capital Latin letters, such as ${\rm I,J,K}$
to represent SO($N$) indices, small Latin letters to represent tensor
indices, and small Greek letters for spinor indices.   Spinors with 
their indices up transform as scalar tensor densities of weight one
(1) while those with their indices down transform as scalar densities of weight
minus one (-1).  For example  the generator $U^{\rm V_1 V_2 V_3}$
has a tensor density realization of contravariant tensor with rank one and
weight $-\frac{3}{2}$ living in the ${\rm N \times  N \times N}$ representation of
SO($N$), i.e. $\o^{{\rm V_1 V_2 V_3}; a}_{\a_1 \a_2 \a_3}$. 
This is the first step at identifying a spectrum of physical fields
that have a natural connection to the algebra.  The completition of this
identification comes from  identifying the tensors related to the
coadjoint representation.  The transformation laws for the coadjoint
elements and tensor representation are tabulated below.  We will raise
and lower the SO($N$) indices using the ${\bf 1}$ matrix, while the
spinor indices are raised and lowered with $C^{\a \b}$ and $C_{\a \b}$.

\begin{center}
\begin{tabular}{|c|c|c|}\hline
\multicolumn{3}{|c|}{ \bf Table 2: Tensors Associated with the Dual of the  Algebra} \\ \hline\hline
${\rm {Dual \,element\, of\, algebra}}$ & ${\rm {Transformation\, Rule}}$ & ${\rm Tensor\, Structure}$ \\ \hline 
$ {L^\star}_{\cal A} ~\to~ D $ &  $ D  ~\to~ -2 \xi' D - \xi D'  ~~~~$
& $ D_{a b}$  \\ \hline
$ {G^\star}^{\rm I}_{\cal A} ~\to~ \psi^{\rm I} $ &  $ \psi^{\rm
  I}~\to~{- \xi(\psi^{\rm I})' - \fracm 32 \xi' \psi^{\rm I}} ~~ $
&$\psi^{\rm I}_{a \a}$  \\ \hline
${T^\star}^{\rm{ R} { S}}~\to~A^{\rm{ R}{ S}} $ &  $ A^{\rm{ R}{ S}}
~\to~  -(\xi)' A^{\rm{ R}{ S}} -\xi\,({A}^{\rm{ R}{S}})'~~~~$ & $ A^{\rm{ R}{ S}}_a $  \\ \hline
${U^\star}^{\rm V_1 \cdots V_n} ~\to~ \o^{\rm V_1 \cdots V_n}  $ &  $
\o^{\rm V_1 \cdots V_n}~\to~  -\xi (\o^{\rm V_1  \cdots  V_n})'  -(2
-\fracm n2) \xi' \o^{\rm V_1 \cdots V_n}~~~~$ &
${\o}^{\rm V_1 \cdots  V_n;\, \a_1 \cdots \a_n}_{a b}  $  \\ \hline
${R^\star}^{\rm T_1 \cdots T_r}_{\cal A}~\to~ \rho^{\rm T_1 \cdots T_r} $ &
${\rho}^{\rm T_1 \cdots T_r} ~\to~   -({\rho}^{\rm T_1 \cdots T_r})'\xi-
(2-\frac r2)  \xi'  \rho^{\rm T_1 \cdots T_r}  ~~~~$ & $ \rho^{{\rm
    T_1 \cdots T_r}; \a_1 \cdots \a_r}_{a b} $ \\ \hline
\end{tabular}
\end{center}
Thus for $N$ supersymmetries there is one rank two tensor $D_{a b}$,
$N$ spin-$\fracm 32$ fields $\psi^{\rm  I},$ 
a  spin-1
covariant tensor $A^{\rm {R}{S}}$ that serves as the $N (N - 1)/2$ SO($N$) gauge 
potentials associated with the supersymmetries, and $N \, (\, 2^N \,-\,
N -1\,)$ fields for both the $ {\o^{\rm {V_1} \cdots {V_p}}}$ fields and $ {\r^{\rm {T_1} \cdots {T_p}}}$
sectors.  

We would like to capture these component fields into superfields and
write an action in terms of these superfields.  One can see from the
above table that at least two distinct superfields will be needed 
to absorb the field content.  These two superfields will constitute a
diffeomorphism sector and a gauge sector for the SO($N$) gauge
symmetry.  

\subsection{The Chiral Diffeomorphism Superfield}
~~~~Instead of using component fields we would like a superfield
formulation of the algebra that can be used to construct an
$N$-extended version of the affirmative action found in
\cite{gates-rodgers}.  Just as in the above reference we will write the
action using tensor notation so that future extension to higher
dimensions and non-chirality can follow easily. Our focus will be on the
two dimensional models, so we will assume in what follows that the
Grassmann variables are Majorana and chiral.  Therefore in this section
we will display a fermionic index,
$''\a''$ say, with the understanding that it is a one dimensional
index.  

 Recall that up to the central extension the Virasoro algebra 
(or Witt algebra) may be realized as the one dimensional reduction of
the Lie algebra of vector fields.  Consider the vector fields $\xi^a$
and $\eta^a$.  We know that the Lie derivative of $\eta^a$ with respect to
${\xi^a}$ is given by
\begin{equation}
{\cal L}_{\xi} \eta^a = -\xi^b \partial_b \eta^a + \eta^b \partial_b
\xi^a 
= (\xi \circ \eta)^a,
\end{equation}
and further that 
\begin{equation}
 [{\cal L}_{\xi}, {\cal L}_{\eta}] = {\cal L}_{\xi \circ \eta}.
\end{equation}
Now consider the superfields ${\cal F} = (\xi^n, \chi^{J,\b})$ and 
${\cal G} = (\eta^n, \psi^{K;\a})$.  We are assuming that we have 
Majorana fermions.  Define the derivative operator ${\cal D}_{I;_\n}$
through
\begin{equation}
 \{ {\cal D}_{I; \m}, {\cal D}_{J; \n}\} = -i \, \d_{I J} \,
\g^n_{\m \n} {\partial \over \partial z^n}. 
\end{equation}
This implies that 
\begin{equation}
 {\cal D}_{I; \m} = {\partial \over \z^{I;\m}} - \d_{I\,J}\,\fracm i2
  \z^{J;\n} \g^m_{\n \m} {\partial \over \partial z^n}.  
\end{equation}
The super Virasoro algebra contains both a diffeomorphism as well as a
supersymmetry transformation with ${\cal D}_{I;\,\m}$.  We can construct vector fields 
\begin{equation} F = \xi^n {\partial \over \partial z^n} + \fracm 12 \chi^{J; \b}
  {\cal D}_{J;\,\b},
\end{equation}
 and 
\begin{equation}
 G = \eta^n {\partial \over \partial z^n} + \fracm 12 \psi^{J; \b}
  {\cal D}_{J;\,\b} .  
\end{equation} 
The commutator of $F$ and $G$ is
{\baselineskip=25pt  
\begin{eqnarray}
[F,G] &=& [ \xi^n {\partial \over \partial z^n} + \fracm 12 \chi^{J; \b}\,
  {\cal D}_{J;\,\b} \, , \, \eta^m \, {\partial \over \partial z^m} \, +
  \, \fracm 12 \psi^{I; \a}\,
  {\cal D}_{I;\,\a} ] \cr
&=& \{ \xi^n \, {\partial \over \partial z^n}\, \eta^m - \eta^m
 \,{\partial \over \partial \, z^m} \, \eta^n + \fracm 12 \chi^{J;\b}\, ({\cal
   D}_{J;\b} \, \eta^m) - \fracm 12 \psi^{K;\a} \, ({\cal D}_{K;\a} \, \xi^m )
\cr
&\,&\,\,\,\,\,\,+ i (\frac12 \chi^{J;\b} )\,(\frac12 \psi^{K; \a})\d^{JK}\,
\g^n_{\b \a}\}\,{\partial \over \partial z^n }\cr
&+& \{ -\fracm 14 (\chi^{J;\b} ({\cal D}_{J; \b} \, \psi^{K;\a} ) +
\psi^{J;\b}( {\cal D}_{J; \b}\, \chi^{K;\a}) \cr
&\,&\,\,\,\,\,\,+\fracm i2 ( \xi^n {\partial
  \over \partial z^n} \psi^{K;\a} - \eta^n {\partial \over \partial
  z^n } \chi^{K;\a})\}\, {\cal D}_{K;\a}\,\,. \label{commutator}
\end{eqnarray}
}
With this in place we can now naturally extend  the vector fields
$\xi^n$ to N supersymmetries.  Let
\begin{equation}
\z^{I_1 \cdots I_m; \a_1 \cdots \a_m } \equiv \z^{[I_1; \a_1}
\cdots \z^{I_m; \a_m]},
\end{equation}
then an $N$-extension of the vector field $\xi^n$ is given by the vector
superfield $F^n$, where 
\begin{equation}
F^n = \xi^n + \chi^{n}_{I_1;\a_1}\, \z^{I_1; \a_1} + r^{n}_{I_1
  I_2 ;\a_1 \a_2}\, \z^{I_1 I_2; \a_1 \a_2} + \cdots +
r^{n}_{I_1 \cdots I_N ;\a_1 \cdots \a_N}\, \z^{I_1 \cdots I_N;
  \a_1\cdots \a_N} 
\end{equation}
This superfield contains the superpartner that is used to perform the
supersymmetric translation.  With this we can write the
superdiffeomorphism vector field that is analogous to the one used in
Eq.[\ref{commutator}] as
\begin{equation}
F = F^n {\partial \over \partial z^n} + \d^{A B} \, ({\cal D}_{A; \,\a}\,
F^n) \, \g^{\a \,\b}_n \, {\cal D}_{B;\,\b}.
\end{equation}
The $F^n$ superfield contains the field content of the $L_{\cal A},
G^I_{\cal A},$ and
$R^{T_1 \cdots T_r}_{\cal A}$ generators seen in Table 1.  Note  that 
$\chi^{I\, \a}  = \delta^{I \, I_1}\,\g^{\a_1\, \a}_n
\chi^{n}_{I_1;\a_1}$. 

However the centrally extended contributions demand that we also
consider the prepotential $F^{\,n\, I_1 \cdots I_N; \a_1 \cdots \a_N}$
given by 
\begin{eqnarray}
 F^{\,n\, I_1 \cdots I_N; \a_1 \cdots \a_N} \, &\equiv & \, 
\xi^n \z^{I_1 \cdots I_N;  \a_1\cdots \a_N} 
+ \chi_{0}^{n\, [I_1 \cdots I_{N-1} ;\a_{N-1}}\, \z^{I_N; \a_N]}\,+\ \cr 
\,&&r_{0}^{n \, [I_1  I_{N-2} ;\a_1 \a_{N-2}}\, \z^{I_{N-1} I_{N-2}; 
\a_{N-1} \a_{N-2}]} + \cdots +
r_{0}^{n I_1 \cdots I_N ;\a_1 \cdots \a_N},
\end{eqnarray} 
where the ``$0$'' subscripted fields are the prepotentials defining
the elements of $F^n$ through
\begin{equation}
F^n = {\cal D}_{ I_1;\,\a_1} \, \cdots \, {\cal  D}_{ I_N;\,\a_N} \, F^{n\,
  I_1 \cdots I_N; \a_1 \cdots \a_N}.
\end{equation}
Then one may write down the commutation relations for centrally
extended elements as
\begin{equation}
[[(F,a), (G,b) ]] = \big([F, G], <<F,G>> \big)
\end{equation}
where the two cocycle $<<F,G>>$ is defined as 
\begin{equation}
<<F,G>>  \equiv  \frac{i c}{24 \p} \int d
\,z \, d\,\z_{I_1 \cdots I_N;\,\a_1\cdots \a_N} \big( (F^n)'''\, G^{m\, I_1  \cdots I_N;\,
  \a_1 \cdots \a_N} - (G^n)''' \, F^{m\, I_1 \cdots I_N; \, \a_1 \cdots
  \a_N} \big) \big).
\end{equation}  In one dimension the tensor indices $m$ and $n$ are not
relevant but simply let us keep track of the transformation properties of the
fields which will be useful in higher dimensions. The integrand transforms as a scalar density in one dimension. This preserves
the $N=0$ form of the two cocycle used for the central extension of
the Virasoro algebra.

To continue we need the dual elements of these vector fields. 
This implies that there exists a bilinear two form $<*|*>$ that is
invariant under diffeomorphisms. Consider the dual of $F^n$ for ${\rm
  N}$ supersymmetries.  In one dimension one has a pairing that can be
represented tensorially as the  integral
\begin{equation} 
<(F, a) | (B, {\hat b})> = \int d{z}^p\, d\z_{\a_1; I_1} \cdots d\z_{\a_N; I_N}  \,
F^n\, B_{n p}^{I_1 \cdots I_N;\,\a_1 \cdots \a_N} + a {\hat b}.
\end{equation}
In one dimension $F^p$ is considered a rank one contravariant vector
field, while $B_{n p}^{I_1 \cdots I_N;\,\a_1 \cdots \a_N}$ is a
  quadratic differential with density $-\frac N 2$. 
In higher dimensions we will take advantage of the fact that $\sqrt{g}
F^{a b}$ transforms like $F^a$ in one coordinate.  The invariant integral in $k$
dimensions will be written as 
\begin{equation} 
<B | F> = \int d{z}^k\, d\z_{\a_1; I_1} \cdots d\z_{\a_N; I_N}  \,
\sqrt{g}\,F^{n p}\, B_{n p}^{I_1 \cdots I_N;\,\a_1 \cdots \a_N}.
\end{equation}  For now, we use the 
one dimensional commutators and pairing to write the transformation law
for the coadjoint elements as  
\begin{eqnarray}
\d_F B_{n p}^{I_1 \cdots I_N;\,\a_1 \cdots \a_N} &=& - F^m\, \partial_m
\,B_{n p}^{I_1 \cdots I_N;\,\a_1 \cdots \a_N} - B_{m p}^{I_1 \cdots
  I_N;\,\a_1 \cdots \a_N} \, \partial_n \, F^m \cr
&-& B_{n m}^{I_1 \cdots
    I_N;\,\a_1 \cdots \a_N}
 \, \partial_p\, F^m + \frac N2 \,(\partial_m
  F^m) \, B_{n p}^{I_1 \cdots I_N;\,\a_1 \cdots \a_N}\cr
&+& {\hat b} c \,\nabla_{n} \nabla_{p} \nabla_{m} F^{m I_1 \cdots I_N;
  \a_1 \cdots \a_N}.
\end{eqnarray}
The transformation shows that $B_{n p}^{I_1 \cdots I_N;\,\a_1 \cdots  \a_N}$
 transforms as a rank two tensor due to its $n$ and $p$
indices and for each contravariant fermion index we have assigned a 
density of weight $-\frac 12$.  

In terms of the fields in Table 2, the field  $B_{n p}^{I_1 \cdots I_N;\,\a_1 \cdots  \a_N}$ has a decomposition of 
\begin{eqnarray}
B_{n p}^{I_1 \cdots I_N;\,\a_1 \cdots \a_N} &=& D_{n p} \, \z^{I_1 \cdots
  I_N;\,\a_1 \cdots \a_N}
+ \psi^{[I_1; \a_1}_{n p} \,\z^{I_2 \cdots I_N;\, \a_2 \cdots \a_N]}\cr
 &+& \r^{[I_1 I_2; \,\a_1 \a_2}_{n p}\, \z^{I_3 \cdots I_N;\, \a_3 \cdots
  \a_N]}
+ \cdots + \r^{I_1 \cdots I_N; \,\a_1 \cdots \a_N}_{n p},
\end{eqnarray}
with the $\frac 32$ spin field $\psi^{I}_{a \a}$ in Table 2
satisfying,
\begin{equation}
\psi^{I}_{a \a} \, = \, \psi^{I; \a_1}_{a p} \,\g^p_{\a_1 \a}\,\,\,.
\end{equation}

In the construction of the actions that follow we will need to define
$F^n$ in terms of the superfield $B_{n p}^{I_1 \cdots I_N;\,\a_1
  \cdots \a_N}$. For the moment let us ignore the SO($N$) gauge symmetry
that has been induced by the $T^{I J}_{\cal M}$ generators. We will use a
tilde as a reminder of this.  Then  we may write 
\begin{equation}
{\tilde B}_{n p} = {\cal D}_{ I_1;\,\a_1} \, \cdots \, {\cal
  D}_{ I_N;\,\a_N} \, B_{n p}^{I_1 \cdots I_N;\,\a_1 \cdots \a_N}.
\label{B}
\end{equation}
Since $F^n$ will generate time-independent coordinate transformations,
we assign $F^n  = {\tilde B}^n_0$.  
Before we put these ingredients into an action we must deal with the
SO($N$) symmetry.  

\subsection{The SO($N$) Gauge Superfield}

~~~~In the previous section we ignored the SO($N$) gauge field that has
become manifest due to the fact that the $T^{I J}$ generators
transform as scalar fields while their dual elements transform as
vector fields under diffeomorphisms.  Again we would like to cast the
algebra into the tidy language of superfields.   
Consider the  superfields $\L^{I; \a}$ that enjoy
the $\z$ expansion in terms of the fields found in Table 1,
\begin{equation}
\L^{ I;\a} = w^{I; \a} + t^{I}_{\,\, J} \z^{J; \a} + w_{I_1 ; \a_1 } \z^{I_1 I ; \a_1 \a} + \, \cdots \, + 
w_{I_1 \cdots  I_{N-1}; \a_1 \cdots \a_{N-1}} \z^{I_1 \cdots I_{N-1}, I ; \a_1
  \cdots \a_{N-1}, \a}\, .
\end{equation}
Here $w^{I; \a} = w^{I ; a}_{\b} \, \g_{a}^{\b \a},$ while $t^{I J}$ is
the anti-symmetric field found in  Table 1. 
A generic element is then  
\begin{equation}
\L=  \L^{J;\a}{\partial \over \partial \z^{J; \a}}. 
\end{equation}
We introduce the algebraic prepotential $\L^{I I_1 \cdots I_N;\, \a \,\a_1
  \cdots \a_N}$ through
\begin{equation}
\L^{J;\,\a} = {\cal D}_{I_1;\,\a_1} \, \cdots \, {\cal D}_{I_N;\,\a_N} \,
\L^{ I [ I_1 \cdots I_N ];\, \a [ \a_1  \cdots \a_N ] }.
\end{equation}
Then the centrally extended commutation relations can be written as
\begin{equation}
[[ (\L, \l), (\O, \m) ]] = ([\L, \O] , << \L, \O >>).
\end{equation}
where the one dimensional two
cocycle for the algebra is 
\begin{eqnarray} 
<<\L, \O>> &=& k\, \int d{z}^p\, d\z_{I_1 \cdots I_N;
  \a_1 \cdots \a_N} \,\big( ({\cal D}_{J;\a}\L^{I;\a})'\, {\cal
  D}_{N;\b}\,\O^{ M\,[ I_1 \cdots I_N];\, \b\,[ a_1 \cdots \a_N]} - \nonumber \\
&&({\cal D}_{J;\a}\O^{I;\a})'\, {\cal  D}_{N;\b}\, \L^{ M\,[ I_1 \cdots I_N];\,
  \b\,[ a_1 \cdots \a_N]}\big) \, \d_{I\,M}^{ J\,N}.
\end{eqnarray}
The $\d_{I\,M}^{ J\,N}$ allows us to show that the SO($N$) invariant trace
and is defined as 
\begin{equation}
\d_{I\,M}^{ J\,N} \, \equiv \, \d_{I M} \,\d^{J N} - \d_{I }^{N}
\, \d^{J}_{ M}.
\end{equation}
  Again notice that the derivative in one dimension has
the affect of changing a scalar into a scalar density.  Thus the
integral is invariant under one dimensional diffeomorphisms.  

Again we would like to extract a dimension independent description
of the dual elements of the SO($N$) adjoint.  Here we use the fact 
that in one dimension $\L^{I;\a}$ transforms the same way as $\sqrt{
g} \L^{I;\,\a p}$.  This device allows us to find a dimension independent
description of the dual of the algebra.   The pairing between a centrally 
extended element, say $(\L^{I;\a}, \, a),$ and a coadjoint element is 
then
\begin{equation}
<(\L,\,a) \mid (A,\b)> =  \int d{z}^k\, d\z_{I_1 \cdots I_N;  \a_1 
\cdots \a_N} \,\sqrt{g}\, \L^{I;\a \,p} \, A^{\,\, \, I_1 \cdots 
I_N; \, \a_1 \cdots \a_N}_{I;\a\, p} + \, a\,\b.
\end{equation}
The dual element has the $\zeta$ decomposition 
\begin{equation}
 A^{\,\, \, I_1 \cdots I_N; \, \a_1 \cdots \a_N}_{I;\a\,p} = \o_{I; 
\a\,p}\, \z^{I_1 \cdots I_N; \, \a_1 \cdots \a_N} + \o_{I;\a\,p}^{[I_1 
; \a_1}\, \z^{I_2 \, \cdots \, I_N;\, a_2 \, \cdots \, a_N]} + \, 
\cdots \,  + \o_{I; \a\,p}^{I_1 \cdots I_N ; \, \a_1 \cdots \a_N}.
\label{vectorpotential1}
\end{equation}
From the algebra and two cocycle we find that the transformation of
the dual element with respect to $\O^{I; \,\a}$ is
\begin{eqnarray}
\d_{\O} A^{\,\, I_1 \cdots I_N; \, \a_1 \cdots \a_N}_{I\,;\a\,p}\,&=&\,
- \O^{J; \b} \,{\cal D}_{J;\b}  A^{\,\,  I_1 \cdots I_N; \, \a_1
\cdots \a_N}_{I\,;\a\,p}\, + ({\cal D}_{I;\b}\, \O^{J; \b})  A^{\,\,  
I_1 \cdots I_N; \, \a_1 \cdots \a_N}_{J\,;\,\a\,p}\,\nonumber \\ &-&
({\cal D}_{J;\b}\,
\O^{J; \b})\, A^{\,\,  I_1 \cdots I_N; \, \a_1 \cdots
  \a_N}_{I\,;\a\,p}\, \nonumber\\
&+ & {\hat b}\,{k}\, {\cal D}_{J;\b}{\cal D}_{N;\a}\nabla_p \O^{
M\,[ I_1 \cdots I_N];\,\b\, [\a_1 \cdots \a_N]}\,\d_{I M}^{~~\,J N}
\, , \end{eqnarray} 
This transformation law is the the Lie derivative on the SO($N$) group
manifold where only the ``I'' index is recognized, since the other
SO($N$) indices were contracted with the SO($N$) volume form.  Notice 
that the coadjoint element transforms as a tensor density of weight $1$
due to the penultimate term in the transformation law.

The gauge field can be extracted from the superpotential by writing
\begin{equation}
A^{\,\,\,I}_{J\,p}= {\cal D}_{I_2;\,\a_2} \cdots {\cal D}_{I_N;\,
\a_N}\,  A^{\,\, \, I\,[I_2 \cdots I_N]; \, \a\,[\a_2 \cdots
  \a_N]}_{J;\,\a\,p}. \label{vectorpotential2}
\end{equation}

We can then build an SO($N$) covariant derivative operator so that 
the covariant derivative on a vector living in the left regular
representation, say
$B^L_q$, is
\begin{equation}
\nabla_p \, B^M_q = \partial_p \, B^M_q - \G_{p q}^{\,\,\,r}\,B^M_r  
+ A^{\,\,\,J}_{I\,p}\, B^L_q \,(\d^{I}_{\,\,L} \, \d_{J}^{\,\,M} - 
\,\d_{J L}\,\d^{I\,M}) 
\label{derivative}
\end{equation}
Then Eq.(\ref{B}) may be written as
\begin{equation}
{B}_{n p} = {\nabla}_{ I_1;\,\a_1} \, \cdots \, {\nabla}_{ I_N;\,\a_N} 
\, B_{n p}^{I_1 \cdots I_N;\,\a_1 \cdots \a_N}. 
\end{equation}

\section{The $N$-Extended Affirmative Action} 

~~~~We are now in a position to construct the affirmative action.  
In \cite{gates-rodgers} the principles used to  construct the $N=1$
action were distilled into a short procedure.  In the $N=0$ case, 
the action is written as  
\begin{eqnarray}
\label{action}
S_{\mbox{\tiny N=0}}=-&\int& d^nx \sqrt{g}~\fracm1q
\left( X^{l m r }~{\rm D}^a{}_r X_{m l a} +2 X^{l m r} 
{\rm D}_{l a} X^a{}_{r m}\right)\\
-&\int& d^nx \sqrt{g}\left(\frac{1}4  X^{a b}{}_b {} 
\nabla_l \nabla_m{} X^{l m}{}_a+ \frac{\b}2  X^{b g a} 
X_{b g a}\right) ~~~, \nonumber \label{diffeq0}
\end{eqnarray}
where $D^{m n}$ is the $N=0$ remnant of the superfield
$B_{m n}^{I_1\cdots I_N; \a_1\cdots \a_N}$ and 
${\rm X}^{m n r} = \nabla^r D^{m n}$.

To the extent possible, we will use the  $N=0$ action as a prototype
to building the $N$-extended affirmative action.    
The construction goes in the following stages:
\begin{enumerate}
\item Contract the conjugate momentum of the field with the 
variation of the field:
\begin{enumerate}
\item $N =0$
\begin{equation}
{\cal L}_{0}={\rm X^{i j 0}\big(\xi^l \partial_l D_{i j}+D_{l j}
\partial_i \xi^l + D_{i l}\partial_j \xi^l \big)} ~~~.
\end{equation}

\item $N$-Extended

The physical fields are contained in the superfield $B_{m n}^{I_1\cdots 
I_N; \a_1\cdots \a_N}$ and the variation is with respect to the superfield
$F^m$.  We use the derivative operator $\nabla_p$ from Eq.(\ref{derivative}) which is covariant with respect to the SO($N$)
symmetry and general coordinate transformations.  The $B_{m n}^{I_1\cdots
I_N; \a_1\cdots \a_N}$ field lives in the direct product space of N left
regular representations. In order to recover the $N =0$ results we write 
\begin{equation}
{\cal L}_{0^{th}}=\nabla_0 B^{\{I\} }_{pq}\,
  \d B^{pq}_{\{I\}}\\ \nonumber  ~~~,
\end{equation}
where 
\begin{equation}
 B^{\{I\} }_{pq} \equiv B_{p q}^{I_1\cdots I_N; \a_1\cdots \a_N}
\end{equation}
and
 \begin{eqnarray}
\d B_{n p}^{I_1 \cdots I_N;\,\a_1 \cdots \a_N} &=& - F^m\, \nabla_m
\,B_{n p}^{I_1 \cdots I_N;\,\a_1 \cdots \a_N} - B_{m p}^{I_1 \cdots
  I_N;\,\a_1 \cdots \a_N} \, \nabla_n \, F^m \cr
&-& B_{n m}^{I_1 \cdots
    I_N;\,\a_1 \cdots \a_N}
 \, \nabla_p\, F^m + \frac N2 \,(\nabla_m
  F^{m}) \, B_{n p}^{I_1 \cdots I_N;\,\a_1 \cdots \a_N}\cr
&+& {\hat b} c \,\nabla_{n} \nabla_{p} \nabla_{m} F^{m I_1 \cdots I_N;
  \a_1 \cdots \a_N}.
\end{eqnarray}

\end{enumerate}

\item Replace the fields $F^m\,(\xi^i)$ with a space-time component of 
the field, $B^m_0 \, (D^a_0)$:
\begin{enumerate}
\item $N =0$ 
\begin{equation}
 {\cal L}_{1^{st}}={\rm X^{L M 0}\big(D_0{}^A 
\nabla_A  D_{L M}+D_{A M}\nabla_L 
D_0{}^A + D_{L A}\nabla_M D_0{}^A \big)} ~~~. \label{DS2}
\end{equation}
\item $N$-Extended 
 \begin{eqnarray}
{\cal L}_{1^{st}} &=&\nabla^0 B_{\{I\}}^ {n p}\,\big( - B_0^m\, \nabla_m
\,B_{n p}^{I_1 \cdots I_N;\,\a_1 \cdots \a_N} - B_{m p}^{I_1 \cdots
  I_N;\,\a_1 \cdots \a_N} \, \nabla_n \, B_0^m \cr
&-& B_{n m}^{I_1 \cdots
    I_N;\,\a_1 \cdots \a_N}
 \, \nabla_p\, B_0^m + \frac N2 \,(\nabla_m
  B^{m}_0) \, B_{n p}^{I_1 \cdots I_N;\,\a_1 \cdots \a_N}\cr
&+& {\hat b} c \,\nabla_{n} \nabla_{p} \nabla_{m} B_0^{m I_1 \cdots I_N;
  \a_1 \cdots \a_N}\big).
\end{eqnarray}
\end{enumerate} 
\item Extend the time directions to covariant directions:
\begin{enumerate}
\item $N =0$
\begin{equation}
 {\cal L}_{\rm int}={\rm X^{L M R}\big(D_R{}^A 
\nabla_A  D_{L M}+D_{A M}\nabla_L D_R{}^A + 
D_{L A}\nabla_M D_R{}^A \big)}~~~. 
\label{KM}
\end{equation}

\item $N$-Extended
 \begin{eqnarray}
{\cal L}_{int} &=&\nabla^q B_{\{I\}}^ {n p}\,\big( - B_q^m\, \nabla_m
\,B_{n p}^{I_1 \cdots I_N;\,\a_1 \cdots \a_N} - B_{m p}^{I_1 \cdots
  I_N;\,\a_1 \cdots \a_N} \, \nabla_n \, B_q^m \cr
&-& B_{n m}^{I_1 \cdots
    I_N;\,\a_1 \cdots \a_N}
 \, \nabla_p\, B_q^m + \frac N2 \,(\nabla_m
  B^{m}_q) \, B_{n p}^{I_1 \cdots I_N;\,\a_1 \cdots \a_N}\cr
&+& {\hat b} c \,\nabla_{n} \nabla_{p} \nabla_{m} B_q^{m I_1 \cdots I_N;
  \a_1 \cdots \a_N}\big)\cr
&=&\nabla_a B^{\{I\} }_{pq}  B^{pqa}_{\{I\} } .
\end{eqnarray}
\end{enumerate}
\end{enumerate} 

At this point one may add the symplectic part of the Lagrangian to the
above and write the $N$-extended superdiffeomorphism contribution to the
action as 
\begin{equation}
S_{\rm N-diff} =\int d^2x d^n\z\,\sqrt{g} \,\{ \frac{1}{q} \nabla_a 
B^{\{I\} }_{pq}  B^{pqa}_{\{I\} } +
   \frac{\b}2 \nabla_a B^{\{I\}}_{pq} \nabla^a B^{pq}_{\{I\}}\},
\end{equation}
where $q= {\hat b}\, c$, the ``charge'' of the superdiffeomorphism field 
times the central extension.

In a  similar way the SO($N$) gauge field has a contribution to the
action given by
\begin{equation}
S_{\rm SO(N)} ={k{\hat b}} \int d^2x d^n\z\,\sqrt{g} \, (F_{q p}^{
\{I\}})^{LM}\,(F^{q p}_{\{I\}})^{CD} \left( \delta_{LC}
\delta_{MD}-\delta_{LD} \delta_{MC}  \right)
\end{equation}
Here 
\begin{equation}
F_{q p}^{ L M \{I\}} =  \partial_{[q}A_{p]}^{L M\{I\}} - {1 \over{k
{\hat
b}}}\,A^{B
A}_{[q}A^{JN\{I\}}_{p]}f_{AB,JN}^{~~~~~~~~~~ LM},
\end{equation}
where the vector potential is defined through
Eqs.(\ref{vectorpotential1}, \ref{vectorpotential2}).  The SO($N$)
structure constants $f^{AB,JN,LM}$ may
be written as 
\begin{equation}
f^{AB,IJ,LM} \,=\, \d^{AI}\,\d^{BL}\,\d^{JM} +
\d^{BL}\,\d^{LA}\,\d^{IM} - \d^{AJ}\,\d^{BL}\,\d^{IM} -
\d^{BI}\,\d^{AL}\,\d^{MJ}.
\end{equation}

There is one more contribution to this action that has to be accounted
for and that is the interaction between the superdiffeomorphism field 
and the gauge field.  In \cite{curto} it was shown that the coadjoint
elements also transform under gauge transformation.  In terms of the
${\cal GR}$ generators one has, for example, that 
\begin{equation}
\d_{t^{I J}} \, D = \frac12 ( t^{I J})'\, A^{L M}
\,(\d_{I L} \,\d_{J M} - \d_{I M} \, \d_{L J})
\end{equation} as well as the corresponding relationship that
\begin{equation}
\d_\xi A^{I J} = \, -\xi (A^{I J})' - (\xi)' A^{I J}.
\end{equation}
We have already seen this second transformation from Table 2, which was
used to show that $A^{I J}$ transforms as a vector under time-independent 
coordinate transformations.  The first relationship is dual to this 
coordinate transformation.  Such dual relationships have been discussed 
at length in \cite{diff2} and \cite{lano2} and constitute new interaction
Lagrangians between the gauge fields and the diffeomorphism fields. 
The vanishing of the transformation laws determines the condition for 
the isotropy algebra.   This condition it realized as a constraint
equation when the adjoint elements are replaced with the conjugate
momenta.  This follows since the conjugate variables transform the 
same way as the adjoint elements under time-independent transformations.  
Therefore a new interaction term between the gauge sector and the
superdiffeomorphism sector must be present.  

In order to illustrate this we recall the result of \cite{diff2}.
There one considers the interactions due to an affine Lie algebra
interacting with the Virasoro algebra.  The coadjoint elements
transform under simultaneous coordinate and gauge transformations as

\begin{equation}
\d_{\cal F}\,{\rm B}=\left( \xi \left( \theta \right) ,\Lambda \left(
\theta \right) ,a\right) *\left( {\rm D}\left( \theta \right) ,{\rm A
}\left( \theta \right) ,\mu \right) =\left( {\rm \d D}\left( \theta 
\right),{\rm \d A}\left( \theta \right),0\right) \mbox{.}
\label{coadjoint} 
\end{equation}
Here ${\cal F}=\left( \xi \left( \theta \right) ,\Lambda \left( 
\theta \right) ,a\right) $ is an arbitrary adjoint element, while
${\rm B=%
}\left( {\rm D}\left( \theta \right) ,{\rm A}\left( \theta \right) ,
\mu \right) $ is the coadjoint element.  The components of the variation
are
\begin{equation}
{\rm \d D}\,=\,2\xi'\,{\rm D}+{\rm D}'\,\xi +\frac{c\mu}{24\pi }
\xi^{\prime \prime \prime }-{\rm Tr}\left({\rm A}\Lambda^{\prime 
}\right)
\end{equation}
and
\begin{equation}
{\rm \d A}\,=\,{\rm A}^{\prime }\xi +\xi^{\prime }{\rm A} - [\Lambda
\,{\rm A}-{\rm A\,}\Lambda ]+k\,\mu \,\Lambda^{\prime }.
\end{equation}
This leads to the interaction Lagrangian, 
\begin{equation}{\rm \sqrt{g}\,F^{r l}
(D^a{}_r\partial_a {\tilde A}_l + {\tilde A}_a \partial_l D^a_r
-\partial_r(D^a{}_l {\tilde A}_a))},\label{gaugenew} \end{equation}
where 
\[ {\rm {\tilde A}}_m = {\rm A}_m -V^{-1}\partial_m V. \]
The $V$ field corresponds to the WZNW field that maintains gauge
invariance in the interaction of the gauge field  with the
diffeomorphism field.  In fact the vacuum expectation value of $V$ in
the two dimensional case in \cite{diff1} is the identity.   We then 
also require a symplectic contribution to the action for this
field and write it as 
\begin{equation}
S_V = m^2_A \int \sqrt{g}\, (V^{-1}\partial_m V - {\rm A}_m)
(V^{-1}\partial_n V -{\rm A}_n)(g^{m n} + D^{m n}) d^n x.
\end{equation}
The coupling constant $m^2_A$ is to suggest a dynamical origin of a
mass for the gauge field about the vacuum expectation value for $V$.

 Using the results of \cite{diff2} we may write the corresponding
interaction term between the superdiffeomorphism superfield and the
SO($N$) super gauge field as 
\begin{eqnarray}
S_{int} =&{{k {\hat b} \over q \b}\,}&\int d^2x d^n\z\,\sqrt{g} \,
  \left( F^{ab}_{\{I\}} \right)^{LM} \left( -B^n_a  \partial_n {\tilde
  A}_m^{A B \{I \} } -\G_{a m}^p  {\tilde A}_p^{A B \{I \} }\right)
\left( \delta_{LA}\delta_{MB}-\delta_{LB} \delta_{MA} \right)\cr
 -&{{k {\hat b} \over q \b}\,}& \int d^2x d^n\z\,\sqrt{g} \,\left( 
F^{ab}_{\{I\}} \right)^{LM}\,\left( {\tilde A}^{A B \{I\}}_n 
\nabla_b B^n_a \right) \left( \delta_{LA}\delta_{MB}-\delta_{LB}
\delta_{MA} \right)\cr
 +&{{k {\hat b} \over q \b}\, }& \int d^2x d^n\z\,\sqrt{g} \ F^{
ab~LM}_{\{I\}} \left( \frac{N}2 \left( \nabla_n B^n_a \right) 
{\tilde A}^{AB\{I\}}_b \right) \cr
 +& m^2_A& \int d^2x d^n\z\,\sqrt{g} \,  {\tilde A}^{AB}_{a\{I\}} 
\left( {\tilde A}^{CD}_{\{I\}} \right)^a \left( \delta^{AC} \delta^{
BD} - \delta^{AD} \delta^{B C} \right).
  \} 
\end{eqnarray}
The superfield ${\tilde A}_p^{I J} = A_p^{I J} - V_p^{I J}$.  It is 
the analog of ${\tilde A}_p$ in the $N=0$ case mentioned above.  The
corresponding WZNW field is defined so that $V_p^{I J} \equiv {\cal
D}_{\a_1}\cdots {\cal D}_{\a_N}\, V_p^{I J  I_1 \cdots I_N; \a_1 \cdots
\a_N}$ is pure gauge.   The full chiral $N$-extended affirmative action
is then 
\begin{eqnarray}
S_{N\chi AA}& =&
\int d^2x d^n\z\,\sqrt{g} \,\{ \frac{1}{q} \nabla_a B^{\{I\} }_{pq}
  B^{pqa}_{\{I\} } +
   \frac{\b}2 \nabla_a B^{\{I\}}_{pq} \nabla^a B^{pq}_{\{I\}}\}\cr
+&{k{\hat b}}& \int d^2x d^n\z\,\sqrt{g} \,   (F_{q p}^{\{I\}})^{
LM}\,(F^{q p}_{\{I\}})^{CD} \left( \delta_{LC} \delta_{MD}-\delta_{LD}
\delta_{MC}  \right)\cr +&{{k {\hat b} \over q \b}\,}&\int d^2x
d^n\z\,\sqrt{g} \,
  \left( F^{ab}_{\{I\}} \right)^{LM} \left( -B^n_a  \partial_n {\tilde
  A}_m^{A B \{I \} } -\G_{a m}^p  {\tilde A}_p^{A B \{I \} }\right)
\left( \delta_{LA}\delta_{MB}-\delta_{LB} \delta_{MA} \right)\cr
 -&{{k {\hat b} \over q \b}\,}& \int d^2x d^n\z\,\sqrt{g} \,\left( 
F^{ab}_{\{I\}} \right)^{LM}\,\left( {\tilde A}^{A B \{I\}}_n
\nabla_b B^n_a \right) 
\left( \delta_{LA}\delta_{MB}-\delta_{LB} \delta_{MA} \right)\cr
 +&{{k {\hat b} \over q \b}\, }& \int d^2x d^n\z\,\sqrt{g} \ F^{
ab~LM}_{\{I\}} \left( \frac{N}2 \left( \nabla_n B^n_a
  \right) {\tilde A}^{AB\{I\}}_b \right) \cr
 +&m^2_A& \int d^2x d^n\z\,\sqrt{g} \,  {\tilde A}^{AB}_{a\{I\}} 
\left( {\tilde A}^{CD}_{\{I\}} \right)^a \left( \delta^{AC} \delta^{
BD} - \delta^{AD} \delta^{B C} \right).
\end{eqnarray}

\section{Conclusions}
~~~~We have constructed an action for two dimensional chiral supergravity
interacting with the $N$-extended Virasoro group using methods developed
in Refs.\cite{diff1,diff2}.  The $N$-extended theory requires two
superfields to carry the physical fields and auxiliary fields through 
a supersymmetric explicit construction.  These two superfields correspond
to the multiplet containing the gravitational fields related to the
dual of the $N$-extended super Virasoro algebra \cite{gates-rana} and
another superfield to support the gauged SO($N$) vector potential that
arises from gauging the $N$ supersymmetry generators.  One finds that 
a new superfield, $V_p^{I J}$, becomes manifest in order to maintain 
the SO($N$) gauge symmetry.   In later work we will examine whether
supersymmetry can be spontaneously broken when this field takes a 
vacuum expectation value.  

The appearance of the superfields $B_{n p}^{I_1 \cdots I_N;\,\a_1 
\cdots \a_N}$ and $A^{~ \, I\,[I_2 \cdots I_N]; \, \a\,[\a_2 \cdots
\a_N]}_{J;\,\a\,p}$ are very suggestive that in a completely geometrical
approach to the description of 2D, ($N$, 0) supergravity, these quantities
could play the role of supergravity prepotentials.  Should such an 
interpretation be viable, then the representation theory of the
supergravity prepotentials in the context of 2D, ($N$, 0) supergravity
will have been shown to be a direct consequence of the geometrical
arbitary $N$ and model-independent realization of the super-Virasoro
algebra.  Thus a future challenge of this line of investigation is
to show that this algebra approach leads to the unconstrained supergravity
variables required for the curved superspace.  If we anticipate success
in these efforts, then a pressing problem becomes their extension first
to non-chiral 2D, ($N$, $N$) supergravity theories and beyond to even
more complicated higher dimensional theories.  If these off-shell theories
can be interpreted as the toric reduction of higher dimensional theories,
then we will have a proof that all unconstrained supergravity theories
are representations of the super-Virasoro algebra.

\eject

\section{Acknowledgments}

~~~~We thank Mary Ridgell and the University of Maryland for coordinating
this research effort. V.G.J.R thanks the Mathematical Sciences Research 
Institute in Berkeley, CA for hospitality.  A.B., B.L., D.M.K. and
V.G.J.R  thank the University of Maryland for hospitality.

\end{document}